\documentclass[aps,pre,onecolumn,groupedaddress,showpacs]{revtex4}

\usepackage{graphicx}

\begin{document}

\title{Long Range Hops and the Pair Annihilation Reaction $A+A
\rightarrow \emptyset$: Renormalization Group and Simulation}

\author{Daniel C.~Vernon} \email[]{dvernon@sfu.ca}
\affiliation{Department of Physics, Simon Fraser University, Burnaby
BC, Canada V5A 1S6}

\date{December 22, 2003}

\begin{abstract}
A simple example of a non-equilibrium system for which fluctuations
are important is a system of particles which diffuse and may
annihilate in pairs on contact.  The renormalization group can be used
to calculate the time dependence of the density of particles, and
provides both an exact value for the exponent governing the decay of
particles and an $\epsilon$-expansion for the amplitude of this power
law.  When the diffusion is anomalous, as when the particles perform
L\'evy flights, the critical dimension depends continuously on the
control parameter for the L\'evy distribution.  The
$\epsilon$-expansion can then become an expansion in a small
parameter.  We present a renormalization group calculation and compare
these results with those of a simulation.
\end{abstract}


\maketitle

Many different approaches have been used to study the dynamics of
systems far from equilibrium~\cite{privman97, hinrichsen00,
tauber02}. These include exact solutions derived by mapping the
problem to a quantum spin system~\cite{schutz00} or by studying
various particle distribution functions~\cite{masser01, privman97},
renormalization group methods~\cite{cardy96}, and simulations.  Many
of these systems can be assigned to a small number of universality
classes, based on the time dependence of a few measurable quantities.
Renormalization group methods are particularly useful in
characterizing this universal behavior, as they make it possible to
examine the action describing the behavior of the system and determine
which scaling variables are relevant.  Systems of particles which can
diffuse and undergo reactions are straightforward examples, which can
be studied far from equilibrium.  One of the simplest examples of a
reaction-diffusion system is the pair-annihilation reaction, in which
members of a single species of particle, denoted by $A$, react at some
rate $\lambda$ to form an inert product.  This reaction is written $A
+ A \rightarrow \emptyset$.  This reaction, and some related ones,
have been studied for some time.  For some examples, see the work of
Smoluchowski~\cite{smoluchowski17}, who studied the coagulation of
colloidal particles unsing a mean-field approach, and Ovchinnikov and
Zeldovich~\cite{ovchinnikov78}, who examined the effects of
fluctuations on the reaction $A+B\rightarrow \emptyset$. The solution to the (mean-field) rate
equation
\begin{equation}
\label{rate}
\frac{\partial n}{\partial t} = D\nabla^2 n -\lambda n^2
\end{equation}
for the density of particles is the correct result for spatial
dimension $d>2$, but for lower dimensions, fluctuations become
important.  If the number density of particles scales as $n\sim
At^{-\alpha}$ for large $t$, then $\alpha=1$ for $d>2$.  A
renormalization group study by Lee~\cite{lee94} produced the exact
result $\alpha=d/2$ for $d<2$, and also yielded the amplitude $A$ as
an expansion in $\epsilon=2-d$.  The agreement between this amplitude
and the exact result~\cite{lushnikov87} for a specific model in $d=1$
was poor, as might be expected as the expansion parameter $\epsilon=1$
is large here.  However, the $\epsilon$-expansion does provide a
systematic picture of the scaling behaviour of this process.

This paper extends the renormalization group calculation
of~\cite{lee94} to the case of anomalous diffusion, which is modelled
by a long-range hopping process in which the distance a particle
travels in each time step is chosen from a L\'evy distribution.  These
distributions form a family that share with the gaussian the property
that they are ``stable,'' in the sense that the probability
distribution of the sum of two numbers chosen from a L\'evy
distribution is the same distribution, up to a trivial rescaling.  The
probability distribution for each of these distributions has a Fourier
transform
\begin{equation}
\label{levydist}
P(k) = e^{-D_Ak^{\sigma}}, \qquad (0< \sigma \le 2),
\end{equation}  
where $\sigma$ is a parameter that controls the shape of the
distribution and $D_A$ scales the distribution.  For $\sigma<2$, the
real-space distributions have power-law tails, $P(r)\sim
r^{-(d+\sigma)}$ for large $r$. These distributions appear in a number
of physical contexts~\cite{klafter96}, including diffusion in
disordered media~\cite{bouchaud90} and motion of particles in
turbulent flow, in both experiments~\cite{solomon93,weeks98}, and
theoretical calculations~\cite{chen01}. Hinrichsen and
Howard~\cite{hinrichsen99} previously simulated the process studied
here, and determined the exponent $\alpha$, but did not perform a
renormalization group calculation and did not measure the amplitude.

The calculation of the density of particles uses reasonably standard
renormalization group techniques, and will only be summarized here.
The density of particles can be found by solving the Langevin equation
\begin{equation}
\label{langevin}
\frac{\partial \phi({\bf x}, t)}{\partial t} = (D_N\nabla^2 +
	D_A\nabla^{\sigma})\phi({\bf x}, t) -\lambda \phi^2({\bf x},
	t) + \phi({\bf x}, t) \zeta({\bf x}, t)
\end{equation}
with an initial condition set by the initial density of particles
$n_0$.  The anomalous diffusion term, $ D_A\nabla^{\sigma}$, is
defined by its action in Fourier space,
\begin{equation}
D_A\nabla^{\sigma} e^{i{\bf k}\cdot{\bf x}} = -D_A|{\bf
  k}|^{\sigma}e^{i{\bf k}\cdot{\bf x}},
\end{equation}
and the noise term $\zeta$ has correlations of the form
\begin{equation}
\label{noise}
\langle \zeta({\bf x}, t) \zeta({\bf x}', t') \rangle =
-2\lambda\delta^d({\bf x}-{\bf x}') \delta(t-t').
\end{equation}
The average density is then given by an average of solutions to
Eq.~(\ref{langevin}) over noise histories, $n=\langle \phi\rangle$.
The normal diffusion term $D_N\nabla^2$ appears if the distribution of
hops has any component proportional to $k^2$ in its Fourier transform.
This term will be dropped, as it is less relevant than the anomalous
term and flows to zero under renormalization.

The form of the noise term and the noise-noise correlation function
given by Eq.~(\ref{noise}) are not determined by equilibrium physics,
but can be derived from the master equation describing the microscopic
behavior of the system, using the procedure developed by
Doi~\cite{doi76} and Peliti~\cite{peliti85}.  Their procedure produces
an effective field theory describing the behavior of the system, with
an action given by
\begin{equation}
\label{action}
S=\int d^dx \left[\int_0^t dt\left\{\hat\phi(\partial_t
  -D_A\nabla^{\sigma})\phi +
  2\lambda\hat\phi\phi^2+\lambda\hat\phi^2\phi^2\right\}
  -n_0\hat\phi(0)\right].
\end{equation}
The field $\hat\phi$ is a response field, which plays the same r\^ole
as that in the Martin-Siggia-Rose approach~\cite{cardybook96}.  This
field theory can then be used to derive a Langevin equation by
integrating out the response field, applying the Martin-Siggia-Rose
approach~\cite{cardybook96} in reverse.  For examples of this
derivation applied to similar problems, see~\cite{cardy96}
and~\cite{vernon01}. Either the field theory or the Langevin equation
can be used to develop a renormalized perturbation expansion for the
density of particles.

\begin{figure}
\includegraphics[width=8.5cm]{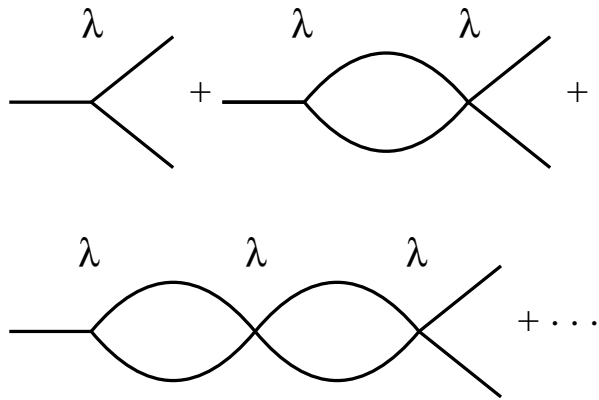}%
\caption{\label{vertex} The diagrams contributing to the renormalized
annihilation rate.  Time flows from right to left.}
\end{figure}

The sum of tree diagrams is equivalent to the solution of
Eq.~(\ref{langevin}) without the noise term.  Divergences appear as
diagrams with loops are included, and these can be handled by
renormalizing the annihilation rate $\lambda$. Power counting shows
that the critical dimension is $d_c=\sigma$, so that the expansion
will be in $\epsilon=\sigma-d$.  The diagrams in Fig.~ \ref{vertex}
give the full renormalization of the annihilation rate, and represent
the sum
\begin{equation}
\lambda_R({\bf k}, t) = \lambda -2\lambda^2\int \frac{{d}
k_1}{(2\pi)^d}\frac{{d} k_2}{(2\pi)^d}(2\pi)^d \delta(k-k_1-k_2)
e^{-k_1^{\sigma}t} e^{-k_2^{\sigma}t}+ ...
\end{equation}

This sum can be done to all orders after a Laplace transformation
$\lambda_R({\bf k}, s)= \int_0^\infty dt e^{-st}\lambda_R({\bf k},
t)$, to give
\begin{equation}
\lambda_R({\bf k}={\bf 0}, s)= \frac{\lambda}{1+
C\Gamma\left(\frac{\epsilon}{\sigma}\right)s^{-\epsilon/\sigma}},
\end{equation}
with
\begin{equation}
  C=\frac{1}{\sigma}\frac{2^{2-d/\sigma}}{(4\pi)^{d/2}}
  \frac{\Gamma(\frac{d}{\sigma})}{\Gamma(\frac{d}{2})},
\end{equation} 
and so, after a renormalization point $s=\kappa^{\sigma}$ is chosen,
the exact flow function for the renormalized annihilation rate $g_R =
\kappa^{-\epsilon}\lambda_R({\bf 0}, \kappa)$ is
\begin{equation}
\beta = \kappa\frac{\partial g_R}{\partial \kappa} = -\epsilon g_R +
\epsilon C \Gamma\left(\frac{\epsilon}{\sigma}\right)g_R^2.
\end{equation}
This has a non-trivial fixed point, which is stable for $d<\sigma$, at
$g_R^*=(C\Gamma(\frac{\epsilon}{\sigma}))^{-1}$.

Imposing the condition that the density be independent of the
(arbitrary) normalization point $\kappa$, and using dimensional
analysis, a renormalization group equation can be written as
\begin{equation}
\left[\sigma D_At\frac{\partial}{\partial(D_At)} +\beta(g_R)
\frac{\partial}{\partial g_R} - n_0 \frac{\partial}{\partial n_0}
+d\right] n_R = 0.
\end{equation}
The solution to this, by the method of characteristics, is
\begin{equation}
n_R(D_At, n_0, g_R,
\kappa)=(\kappa^{\sigma}D_At)^{-d/\sigma}n_R(\kappa^{-\sigma},
(\kappa^{\sigma}D_At)^{d/\sigma},\tilde{g}_R, \kappa)
\end{equation}
where $\tilde{g}_R$ is the running coupling, which goes to $g_R^*$ as
$t\rightarrow\infty$.  In a diagrammatic expansion, each loop brings
in a higher power of the renormalized coupling, so an expansion in the
number of loops is an expansion in $g_R^*$, which is small near the
critical dimension $d_c=\sigma$.  The first approximation to the right
hand side can be found by summing all tree diagrams generated by
expanding either the action (Eq.~(\ref{action})) or the Langevin
equation (Eq.~(\ref{langevin})). The next term, including all diagrams
with loops, is calculated by writing an integral equation for the
density, as done in~\cite{lee94}.  The leading contribution to the
density at long times is then given by
\begin{equation}
n(t) = A(D_At)^{-\frac{d}{\sigma}}.
\end{equation}
The exponent is exact, and, to order $\epsilon^0$, the amplitude is
\begin{equation}
\label{epsilon_expansion}
A=\frac{1}{(4\pi)^{d/2}} \frac{\Gamma\left(\frac{d}{\sigma}\right)}
{{\Gamma\left(\frac{d}{2}\right)}} \frac{2^{1-d/\sigma}}{\sigma}
\left[\frac{\sigma}{\epsilon}-\frac{5}{2}\right].
\end{equation}
This expression is obtained by expanding factors which are singular as
$\epsilon\rightarrow 0$ in $\epsilon$, and leaving the remainder as an
expression in $d$.

\begin{figure}
\includegraphics[width=8.5cm]{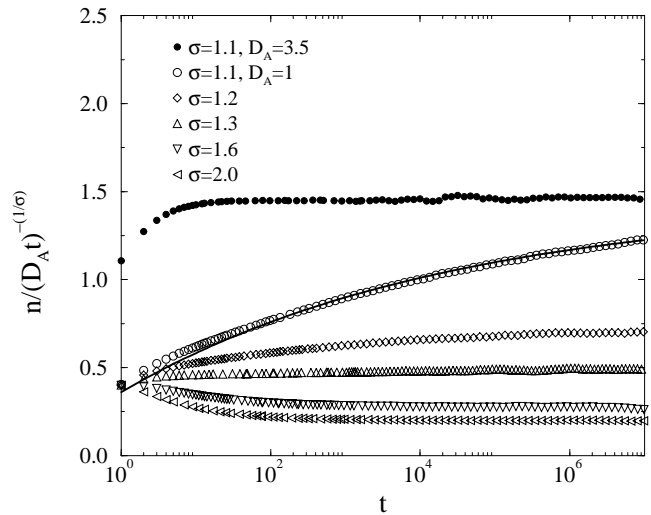}%
\caption{\label{figure:density} The density of particles divided by
its asymptotic (large time) power law, for several values of $\sigma$.
For $\sigma$ close to $1$ and $D_A=1$ (open symbols), the crossover
becomes very slow, but the trend is towards the same value as for
simulations done with a larger value of $D_A$.  The filled symbols
give the density for $D_A=3.5$. Also shown (solid line) is the
crossover function of Eq.~(\ref{crossover}) for $\sigma=1.1$, with the
parameters extrapolated to their $t\rightarrow\infty$ values.}
\end{figure}

Simulations of a microscopic model in one dimension have been
performed and the results compared to the prediction of the
renormalization group calculation.  The convergence of the densities
to the predicted power law is clearly shown in
Fig.~\ref{figure:density}.

The simulations are of a one-dimensional lattice, where only one
particle may occupy each site.  This differs from the renormalization
group calculation, in which multiple occupancy is allowed, but the
annihilation rate flows to a fixed point and thus the bare
annihilation rate does not appear in the final answer, so the results
should be similar.  At the beginning of the simulation, the lattice
contains $L=10^7$ sites, with every site occupied. Whenever the number
of particles fell below 1000, the system was doubled by appending an
exact copy of the current configuration of particles.  While this does
make the system momentarily periodic, the two halves subsequently
evolve differently.  This allowed the simulation to continue to large
times without large statistical fluctuations in the density. For
$\sigma=1.05$, the final system size was $L=2^9\times 10^7$. The
distribution of jumps was chosen to follow a L\'evy stable law, using
the method given in~\cite{bratley87}.  Since this method is for a
continuous distribution, which is then made into a discrete
distribution by rounding, the values generated were multiplied by a
numerically determined factor to produce the correct low-$k$ value of
the distribution.  Earlier simulations~\cite{vernon01, hinrichsen99}
used a pure power-law form for the distribution of jumps, which does
flow to the desired distribution in the long-time limit.  The
distribution used here matches the L\'evy distribution much more
closely after a single step, and makes it easier to determine the
anomalous diffusion constant $D_A$.

For some values of $\sigma$, the anomalous diffusion constant $D_A$ in
the distribution of hop lengths, Eq.~(\ref{levydist}), was varied.  In
the field theory or the Langevin equation, this change to the kinetics
results in a change in the coefficients of powers of $k$, which come
from an expansion of the hop length distribution. These terms, many of
which are irrelevant in the renormalization group sense and are not
written in the Langevin equation, can be seen to have a significant
impact on the crossover behavior, and it may be useful to vary these
parameters in future simulations, to examine the crossover.

As can be seen in Fig.~\ref{figure:density}, the time taken to reach
the asymptotic form of the density depends strongly on $D_A$.  For
$\sigma$ close to $1$ and $D_A=1$, the crossover is very long, but the
amplitude can be extracted from a fit to the form
\begin{equation}
\label{crossover}
\frac{n(t)}{(D_At)^{-1/\sigma}} =A(1-Bt^{-\phi}).
\end{equation}
To obtain the long-time value of the amplitude, the fit was done over
many ranges with differing starting times, and extrapolated to
$t\rightarrow\infty$. For larger values of $D_A$, this fit is not
necessary, as the density reaches its asymptotic value quickly, and
several decades of scaling can be seen in the data.

\begin{figure}
\includegraphics[width=8.5cm]{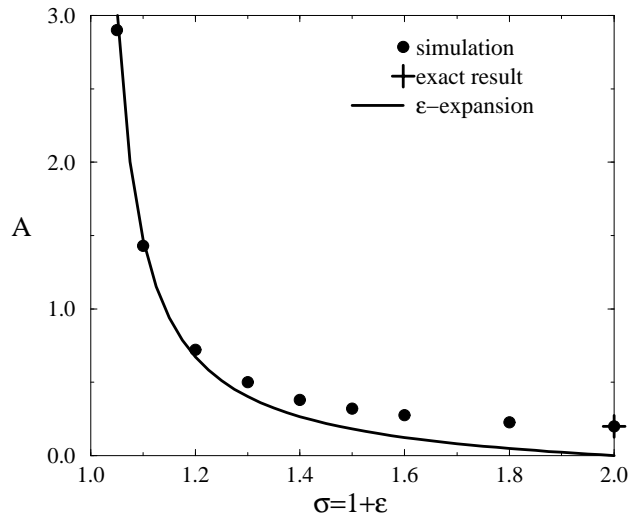}%
\caption{\label{figure:amplitude} The amplitude of the power law decay
of the density of particles determined in simulation (circles),
compared with the renormalization group prediction of
Eq.~(\ref{epsilon_expansion}).  Also shown is the exact
result~\cite{lushnikov87} for the normal diffusion case (cross).  }
\end{figure}

The prediction for the amplitude is compared to simulation results in
Fig.~\ref{figure:amplitude}.  The agreement between the
$\epsilon$-expansion for the amplitude and the simulation result
becomes quite good for small $\epsilon$, as expected for this
asymptotic power series expansion.  As well as being interesting as a
model for physical processes with anomalous diffusion, the L\'evy
flights used here allow this regime, where the expansion parameter is
small, to be explored in a simulation.

\begin{acknowledgments}
Thanks to Michael Plischke for several useful suggestions, and to
Martin Howard and Tibor Antal for helpful comments on an earlier
version of this paper.  Financial support was provided by the NSERC of
Canada.

\end{acknowledgments}


\begin{thebibliography}{20}

\bibitem{privman97}\emph{Nonequilibrium Statistical  Mechanics in One
  Dimension}, edited by V.~Privman (Cambridge University Press,
  Cambridge, 1997).
  
\bibitem{hinrichsen00} H.~Hinrichsen, Advances in Physics \textbf{49},
  815 (2000), \eprint{cond-mat/0001070}.
  
\bibitem{tauber02} U.~C.~Tauber, Acta Physica Slovaca \textbf{52}, 505
  (2002), \eprint{cond-mat/0205327}.

\bibitem{schutz00} G.~M.~Sch\"utz, in \emph{Phase Transitions and
  Critical Phenomena}, edited by {C.}~Domb and J.~L.~Lebowitz
  (Academic Press, 2000), vol.~19.

\bibitem{masser01} T.~O.~Masser and D.~ben-Avraham, Phys.\ Rev.\ E
  \textbf{64}, 066108 (2001), \eprint{cond-mat/0101212}.

\bibitem{cardy96} J.~Cardy, in \emph{The Mathematical Beauty of
  Physics}, edited by J.~M.~Drouffe and J.~B.~Zuber (World Scientific,
  1996), \eprint{cond-mat/9607163}.
  
\bibitem{smoluchowski17} M.~v.~Smoluchowski, Z.\ Phys.\ Chem.\
  (Leipzig) \textbf{92}, 129 (1917).

\bibitem{ovchinnikov78} A.~A.~Ovchinnikov and Y.~B.~Zeldovich, Chem.\
 Phys.\ \textbf{28}, 215 (1978).

\bibitem{lee94} B.~P.~Lee, J.\ Phys.\ A: Math.\ Gen. \textbf{27}, 2633
  (1994).
  
\bibitem{lushnikov87} A.~A.~Lushnikov, Phys.\ Lett.\ A
  \textbf{120},135 (1987).
  
\bibitem{klafter96} J.~Klafter,M.~F.~Shlesinger, and G.~Zumofen,
  Physics Today \textbf{49} (2), 33 (1996).
  
\bibitem{bouchaud90} J.-P.~Bouchaud and A.~Georges, Physics Reports
  \textbf{195}, 127 (1990).
  
\bibitem{solomon93} T.~H.~Solomon, E.~R.~Weeks, and H.~L.~Swinney, PRL
  \textbf{71}, 3975 (1993).

\bibitem{weeks98} E.~R.~Weeks and H.~L.~Swinney, Nonlinear Science
  Today(1998).
  
\bibitem{chen01} L.~Chen and M.~W.~Deem, Phys.\ Rev.\ E
  \textbf{65},011109 (2002), \eprint{cond-mat/0107158}.
  
\bibitem{hinrichsen99} H.~Hinrichsen and M.~Howard, Eur.\ Phys.\ J.\ B
  \textbf{7}, 635 (1999), \eprint{cond-mat/9809005}.

\bibitem{doi76} M.~Doi, J.~ Phys.~A \textbf{9}, 1465 (1976),
  J.~Phys.~A \textbf{9}, 1479 (1976).
  
\bibitem{peliti85} L.~Peliti, J.~Phys.\ (France) \textbf{46}, 1469 (1985).
 
\bibitem{cardybook96} J.~Cardy,\emph{Scaling and Renormalization in
  Statistical Physics}, no.~5 in Cambridge Lecture Notes in Physics
  (Cambridge University Press, Cambridge UK, (1996).
  
\bibitem{vernon01} D.~Vernon and M.~Howard, Phys.\ Rev.\ E
  \textbf{63}, 041116 (2001), \eprint{cond-mat/0011475}.
  
\bibitem{bratley87} P.~Bratley, B.~L.~Fox, and L.~E.~Schrage, \emph{A
  Guide to Simulation} 2nd ed. (Springer-Verlag, New York, 1987).
  
\end{thebibliography}
\end{document}